\documentclass[twocolumn,prl,showpacs,preprintnumbers,amsmath,amssymb,superscriptaddress]{revtex4}

\usepackage{graphicx}
\usepackage{dcolumn}
\usepackage{bm}

\begin{document}


\title{Neutron and Proton Transverse Emission Ratio Measurements and the 
Density Dependence of the Asymmetry Term of the Nuclear Equation of State}
\author{M. A. Famiano}
\email[]{michael.famiano@wmich.edu}
\homepage[]{http://tesla.physics.wmich.edu}
\affiliation{Physics Department, Western Michigan University, Kalamazoo, MI, USA}
\author{T. Liu}
\author{W. G. Lynch}
\author{M. Mocko}
\author{A.M. Rogers}
\author{M.B. Tsang}
\author{M.S. Wallace}
\affiliation{National Superconducting Cyclotron Laboratory, East Lansing, MI, USA}
\author{R.J. Charity}
\author{S. Komarov}
\author{D. G. Sarantites}
\author{L. G. Sobotka}
\affiliation{Washington University in St. Louis, St. Louis, MO, USA}
\author{G. Verde}
\affiliation{INFN, Catania, Italy}
\affiliation{GANIL, CAEN, France}
\date{\today}
\begin{abstract}
Recent measurements of pre-equilibrium neutron and proton transverse  emission from $^{112,124}$Sn+$^{112,124}$Sn
reactions at 50 MeV/A have been completed at the National Superconducting Cyclotron Laboratory.
Free nucleon transverse  emission ratios are compared to those of A=3 mirror nuclei.  Comparisons are made to
BUU transport calculations and conclusions concerning the density dependence of the asymmetry term of the nuclear
equation-of-state at sub-nuclear densities are made. 
The double-ratio of neutron-proton ratios between two reactions is employed as a means of reducing 
first-order Coulomb effects and detector efficiency effects.  Comparison to BUU model predictions indicate a density dependence of
the asymmetry energy that is closer to a form in which the asymmety energy increases as the square root of the
density for the density region studied.  A coalescent-invariant analysis is introduced as a means of
reducing suggested difficulties with cluster emission in total nucleon emission.  Future 
experimentation is presented.  
\end{abstract}


\maketitle

The nuclear symmetry energy increases the masses of nuclei with very different neutron and proton concentrations and limits the neutron concentration and 
maximum neutron number N of any element.  In the interior regions of neutron stars, where neutrons may comprise over 90\% of the matter, 
the symmetry energy may contribute the bulk of the pressure supporting the star \cite{lattimer}. The nuclear equation-of-state at 
densities of 0.5$\le\frac{\rho}{\rho_0}\le$10 (where $\rho_0$ is the nuclear saturation density)
governs many of the neutron star macroscopic properties, including 
radius, moment of inertia, core structure \cite{lattimer_bh}, cooling rates, and the possible collapse of a neutron star into a black 
hole \cite{shen02,page04,steiner05}. 

Constraints on the symmetry energy at sub-saturation density have been obtained from measurements of the diffusion of neutrons and 
protons between nuclei of different asymmetry $\delta$ = (N-Z)/(N+Z) in peripheral collisions \cite{liu04,tsang_liu04}.  
(Here N and Z are the relevant neutron and proton number of the nuclei.)  Measurements of nuclear masses, isovector collective excitations 
and neutron skin measurements may also provide constraints at densities less than $\rho_0$. Nevertheless, the 
density dependence of the symmetry energy is not known well enough to constrain the relevant neutron star properties. Using an alternative 
approach to improve present constraints, new measurements of the ratios of neutron and proton spectra in central heavy ion collisions are 
presented and compared to transport theory calculations. These calculations display a strong sensitivity to the density dependence of the 
symmetry energy, from which additional contraints may ultimately be derived \cite{li}. 
The present study was motivated by an isospin dependent Boltzmann-Uehling-Uhlenbeck (BUU) transport model calculations of Ref. \onlinecite{li}  for 
$^{112}$Sn+$^{112}$Sn and $^{124}$Sn+$^{124}$Sn 
collisions.  Using proton and neutron densities calculated from the
non-linear relativistic mean-field theories as inputs, the dynamics of
nucleon-nucleon collisions are calcuated.  These
calculations utilize nucleonic mean field potentials corresponding to
an equation-of-state 
that can be expressed 
(at zero temperature) in terms of the mean energy of a nucleon in nuclear matter consisting of a sum of terms from a symmetric part and an asymmetric part:
\begin{equation}
E\left(\rho,\delta\right)=E\left(\rho,\delta=0\right)+S\left(\rho\right)\delta^2
\label{symmetry_energy}
\end{equation} 			
where the mean-field component of the symmetry energy is given as a product of the square of the local asymmetry and a 
density dependent factor S($\rho$), which is 
sometimes described at sub-saturation density by a power law S($\rho$) = Cu$^\gamma$ where u$\equiv\frac{\rho}{\rho_0}$.  The uncertainty in
$\gamma$  (0.5$\le\gamma\le$1.6) reflects uncertainties in the nucleon
effective interaction that require better experimental constraints
\cite{parabolic}.
Possible density dependences of the asymmetry term of the EOS -
defined to be the relationship between the density and the energy (or
pressure, 
P=$\rho^2d(E/A)/d\rho$ at constant entropy)
range from ``asy-stiff'' expressions in which the energy and pressure increases more sharply as a function
of density, ``asy-soft'' in which the pressure does not increase as sharply.

Calculations of neutron and proton spectra, for momentum independent mean field potentials with  $\gamma$=0.5 
and $\gamma\approx$1.6  reveal a strong sensitivity of the neutron-proton spectral 
ratio R$_{n/p}$(E$_{cm}$) = (dY$_n$(E$_{cm}$)/dE$_{cm}$) / (dY$_{p}$(E$_{cm}$)/dE$_{cm}$) of transversely emitted nucleons 
to the density dependence of the asymmetry energy, where E$_{cm}$ is
the particle energy in the center-of-mass frame. 
For a strongly density-dependent - or ``asy-stiff'' - system (with
$\gamma\approx$1.6) at sub-saturation densities, the ratio changes
little with 
system asymmetry $\delta$
due to the weakness of the potential at u$<$1 compared to the Coulomb interaction.  However, for a weakly density-dependent - or 
``asy-soft'' - system (with $\gamma$=0.5), R$_{n/p}$(E$_{cm}$) depends
strongly on $\delta$, and a larger emission ratio is found 
for the $^{124}$Sn+$^{124}$Sn reaction
than for the $^{112}$Sn+$^{112}$Sn reaction, reflecting the 
preferential expulsion of neutrons from a neutron rich system due to
the symmetry 
energy \cite{li}.  

Beams of $^{112}$Sn and $^{124}$Sn 
at E/A=50MeV were produced in the K1200 cyclotron at the National Superconducting Cyclotron Laboratory (NSCL).
These beams were incident on $^{112}$Sn and $^{124}$Sn targets located within a thin-walled
aluminum target chamber.  The chamber contained a central charged-particle multiplicity detector, 
a fast plastic scintillator array at forward angles, and three charged-particle telescopes.  Two large area neutron walls
were place outside the chamber.  Each wall had active areas of about 2 m square with each consisting of 
25 horizontally stacked 2 m long quartz glass tubes filled with NE213 liquid scintillator \cite{neutron_wall}.  One wall was placed at a distance of 
4.59 m covering forward polar angles of 8$^0<\theta_{lab}<$34$^{0}$, 
and the other wall was placed at a distance of 5.79 m, covering forward polar angles of 42$^{0}<\theta_{lab}<$62$^{0}$.  
Associated pulse shape discrimination techniques \cite{neutron_wall} were used to distinguish neutrons from $\gamma$-rays. 
During the experiment, shadow bars were inserted between the scattering chamber and the neutron walls to assess the background 
yield of neutrons scattered from the floor and walls of the experimental vault.  Neutron and proton
emission was observed at angles between 70$^0$ and 110$^0$ in the CM to suppress the contribution from decays of the projectile-like fragment (PLF). 

Neutron energies were measured by time of flight relative to the trigger from a segmented BC404 start scintillator array that was placed inside 
the chamber at a distance of 10 cm downstream from the target. 
This array consists of four 3 mm thick plastic paddles, each trapezoidal in shape, arranged at forward angles covering an angular 
region of 5$^0$ to 60$^0$. Each paddle in the array provided an azimuthal angular coverage of 72$^0$. 
A ``missing'' fifth paddle provided a clear flight path from the target to the charged-particle detector array described below. 
The time resolution of the plastic scintillator array was less than 300 ps FWHM and the neutron array time resolution was measured to be about 700 ps FWHM.

Isotopically resolved charged particles with 1$\le$Z$<$3 were measured using three telescopes of the Large-Area Silicon-Strip Array (LASSA) \cite{lassa}.
Each telescope consists of a 500 $\mu$m double-sided silicon strip detector (DSSD) followed by four 6 cm thick CsI(Tl) scintillator detectors arranged in 
quadrants.  Each side of the DSSD consists of 16 strips per side with a strip pitch of 3 mm. The orthogonal strips on the front and back sides of 
each strip detector subdivided the 25 cm$^2$ square surface area into a Cartesian grid of 256 pixels. The LASSA telescopes were arranged in the horizontal 
plane opposite the neutron array at a distance of 20 cm from the target for a total polar angle coverage ranging from 
15$^0<\theta_{lab}<$60$^0$. These telescopes were calibrated with an accuracy of about 
4\% using recoil protons scattered from a CH$_2$ target by an $^{18}$O beam with E/A=50 MeV.

Central collisions were isolated using overall multiplicity as an indicator of the impact parameter \cite{ipp}. 
The impact parameter was measured using the Washington University MicroBall \cite{microball}.  
In its standard configuration, the MicroBall consists of 95 closely packed CsI(Tl) scintillators arranged in
9 rings, covering a polar angle of 4$^0$ to 172$^0$ providing about 97\% of 4$\pi$ coverage.  Scintillation light is collected by 
silicon photodiodes, and particle identification for $^{1,2,3}$H, $^{3,4}$He, Li, Be, and B is accomplished using associated pulse-shape 
discrimination electronics.   For this particular experiment, charged particles emitted at polar angles of 
60$^0\le\theta\le$172$^0$ were detected in 55 elements of the partial MicroBall  array. 
Assuming the charged particle multiplicity  decreases monotonically with impact parameter, a ``reduced'' impact parameter  $\hat{b}$=b/b$_{max}$ 
was determined following Ref. \onlinecite{ipp};  here, events with a
charged-particle multiplicity of N$_C\ge$7 corresponding 
to $\hat{b}\le$0.2 were selected.  
By comparison to simulations of the same collisions with the Anti-Symmetrized Molecular Dynamics (AMD) computer code \cite{ono98}, 
we assessed that this data set corresponded to events uniformly
distributed over impact parameters of b$<$5 fm.  Comparison to the AMD
is made becuase the BUU code does not accurately accomodate
coalescence, and hence, the accuracy of the charged-particle multiplicity at lower
energies suffers.
 
Neutron-proton transverse emission ratios in the CM were compared to predictions given in Ref. \onlinecite{li} for the two systems -
$^{124}$Sn+$^{124}$Sn ($\delta$=0.19) and $^{112}$Sn+$^{112}$Sn ($\delta$=0.11).  These ratios are predicted
to vary with $\delta$ according to the stiffness of the EOS. For a relatively asy-soft EOS, 
proton emission is supressed relative to the Coulomb repulsion at the sub-nuclear densities studied, while proton emission becomes
comparable to neutron emission for a relatively asy-stiff EOS.  A single measure of the stiffness of the asymmetry term is given by the 
double ratio $_1$R$_{124}$/$_1$R$_{112}\equiv$(dn$_n$/dn$_p$)$_{124}$/(dn$_n$/dn$_p$)$_{112}$ of A=1 products for the two 
reactions indicated by the mass of the projectile (and target) in the subscript \cite{double_ratio}.  The double ratio is useful in that
- to first order - detector efficiencies, Coulomb effects, and systematic errors  cancel out, increasing the sensitivity over that of the single
ratio to the symmetry energy.   Ratios of the nuclei $^3$H-$^3$He are also studied since these are predicted to be sensitive to the density dependence of
the nuclear asymmetry term \cite{chen04}.   
\begin{figure}
\includegraphics[width=9.0cm]{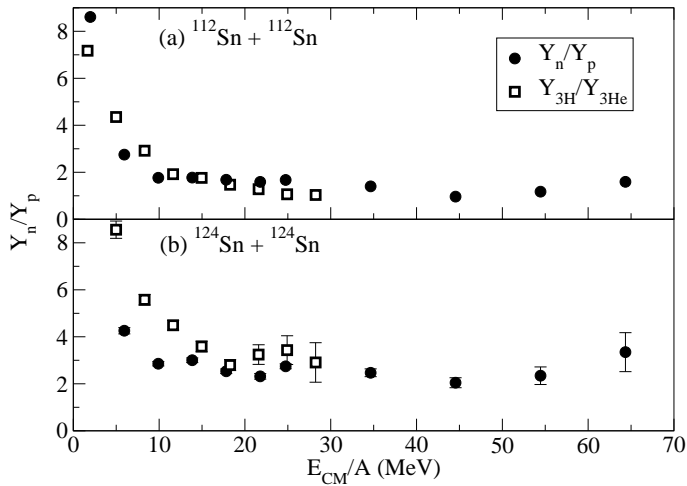}%
\caption{\label{ratios}Ratios of t/$^3$He and n/p for the $^{112}$Sn+$^{112}$Sn reaction (a) and the $^{124}$Sn+$^{124}$Sn reaction (b) as a
function of energy per emitted nucleon for particles emitted between 70$^0$ and 110$^0$ in the center-of-mass.  In the top plot, error bars are
smaller than the points.}
\end{figure}
\begin{figure}
\includegraphics[width=9.0cm]{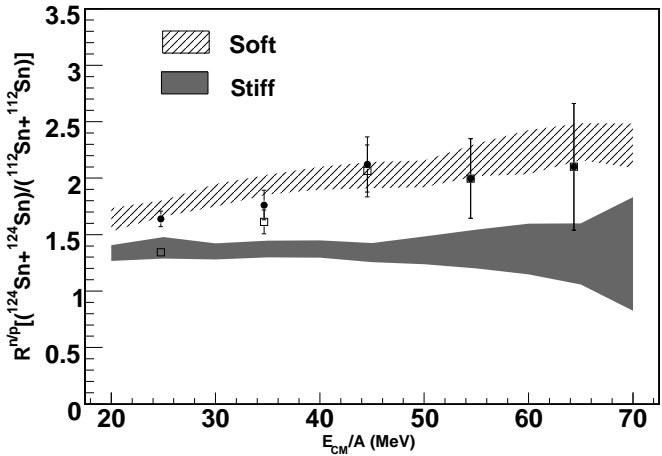}
\caption{\label{double} Double ratios $_1$R$_{124}$/$_1$R$_{112}$  for free nucleons emitted in each reaction compared to the  double ratios
of free nucleon yields calculated from the BUU transport calculations of Ref. \onlinecite{li}.  The filled circles correspond to
double ratios of yields of tranversely emitted 
free nucleons and the open squares correspond to all nucleons including those bound in clusters.}
\end{figure}
Neutron and charged particle energy spectra were obtained for energies up to about 100 MeV in the CM.
Neutron-proton and $^3$H-$^3$He transverse emission ratios are shown in Figure \ref{ratios} for the two reactions studied.
It is noted that the ratios corresponding to the A=3 mirror nuclei are very similar to those of  free nucleons above 20 MeV/A 
with the most discrepancy at low energies where binding energy and
Coulomb effects become considerable.  The A=3 mirror nuclei ratios are
within 10\% of those obtained in recent experiments performed 
by Liu et al.  using the same reactions at the same energy \cite{liu04}.  
Both ratios are lower in the $^{112}$Sn+$^{112}$Sn
reaction than in the $^{124}$Sn+$^{124}$Sn reaction.  
Qualitatively, the results presented in Figure \ref{ratios} 
do not necessarily indicate an  asy-soft EOS, as one naturally expects a higher neutron-proton ratio 
from the $^{124}$Sn+$^{124}$Sn system.  However, given the similarity of the ratios at higher center-of-mass energies, 
this provides evidence that mirror nuclei transverse emission 
ratios may provide a good measure of the density dependence of the asymmetry energy.

Figure \ref{ratios} presents results from an analysis in which cluster emission and free nucleon emission are treated independently. 
However, cluster production may be 
less sensitive to the asymmetry potential for higher energy clusters \cite{chen04}.  More importantly for this analysis, 
the effects of evaporation
from light clusters and the PLF as well as effects from coalescence may not be well understood in the transport calculation. 
In order to explore the effects of cluster yields from fragment decays, 
a coalescent-invariant analysis has been conducted in which isospin emission ratios for a reaction $\Omega$
are taken as the ratios of all 
emitted nucleons - both free and bound:
\begin{equation}
R_{\Omega}=\frac{Y_n}{Y_p}=\frac{\sum_{i}N_iY_i}{\sum_{i}Z_iY_i}
\label{ratio_eq}
\end{equation} 
where the summation is over all species $i$ with charge Z$_i$, neutron
number N$_i$, and yield Y$_i$ up to mass A=16 and charge Z=6.  Because the current experiment only measured 
complete energy spectra over the entire range of particle energies for nuclei with Z$\le$2, 
the fragment yields for this analysis were taken from the
experiment of Liu et al. in which the identical reactions were studied and fragment spectra for Z$>$2 were measured \cite{liu04}.

In order to compare the nucleon transverse emission ratios from each reaction, the double ratio was used \cite{double_ratio}.
The double ratios for free neutrons and protons are shown in Figure \ref{double}.   Also shown are the double ratios calculated from a BUU transport 
calculation for two asymmetry relationships.  In this calculation, a nucleon is considered free if it is not correlated with another nucleon within
a spatial distance and a momentum distance of 3 fm and 300 MeV/c respectively at a time of 200 fm/c after the initial contact of the two reacting
nuclei \cite{li}.  The BUU results correspond to an asy-stiff relationship, with a bulk symmetry energy defined by 
$S{\propto}2u^2/(u+1){\approx}u^{1.6}$, (shown in the shaded region) and 
an asy-soft relationship with a bulk symmetry energy $S{\propto}u^{1/2}$ (shown in the cross-hatched region).  The BUU results shown are for 
central collisions with b$<$5 fm, which is identical to the impact
parameter for the shown experimental results.  Relevant nucleon
emission is only displayed for 
energies
above 20 MeV/A as emission at lower energies is dominated by cluster
emission and is not well modeled by BUU calculations.  Coulomb effects
at low energies can 
also adversely
affect the comparison.  

Care was taken to eliminate some of the experimental uncertainties, as well as those associated with the BUU calculation.
First, because BUU calculations do not include evaporation (while the experiment necessarily does), nucleon emission from the PLF is not included in the
theoretical results.  This is overcome by examining transverse nucleon
emission, because PLF emission is primarily at forward angles.
Second, effects from 
coalescence
are examined via the coalescent invariant 
analysis of Equation \ref{ratio_eq}, in which all nucleons - free and
bound -  are counted in the ratios.  The results of this analysis are
shown in 
Figure \ref{double}.  
Because the results of the coalescent-invariant analysis closely match those of free nucleons at energies above 25 MeV, 
it is concluded that the transverse emission at higher energies is entirely free nucleon emission, and effects from coalescence become negligible.  
The only remaining uncertainty is due to the
fact that no angular cuts are made in the BUU calculation due to limited statistics.  Currently,  BUU calculations are planned  to improve the statistics
near 90$^0$ in the center-of-mass so that such angular cuts will be
possible.  These will be available in the future, and preliminary
results indicate that the 
inclusion of
angular cuts has little effect on the overall results \cite{li_disc}.

The experimentally determined double ratio (the ratio of Y$_n$/Y$_p$
for the $^{124}$Sn+$^{124}$Sn reaction to that for the
$^{112}$Sn+$^{112}$Sn reaction), which may be understood as the change in the proton repulsion (relative to
the neutron repulsion) 
with system asymmetry, appears to correspond more closely to an
asymmetry potential in which $u<F(u)<u^{1/2}$.  In these results, the
relative proton emission with respect to
to the neutron emission increases by over a factor of two as the system asymmetry goes from 0.11 ($^{112}$Sn+$^{112}$Sn) to 0.19 ($^{124}$Sn+$^{124}$Sn).
This can be understood by considering the proton and neutron mean field potentials corresponding to strongly and weakly density dependent asymmetry energies 
at sub-saturation densities.
For the densities less than $\rho_0$ studied here the proton (neutron) attractive (repulsive) potential for an asy-soft EOS increases more sharply with density 
than for an
asy-stiff EOS.   The net result is that the proton attraction competes more with the Coulomb repulsion in the asy-soft case as the system asymmetry changes, 
producing the larger double ratio shown in Figure \ref{double} due to the repressed proton emission and the enhanced neutron emission, compared to the asy-stiff case 
in which the proton emission changes little as the asymmetry changes.  
These results are consistent with predictions of the $^{208}$Pb skin thickness and with
predictions from isospin diffusion, in which the deduced asymmetry energy $S$ is roughly proportional to $u^{0.69}$ \cite{li_chen05,steiner05b}.

Nucleon transverse emission has been studied as a sensitive observable of the asymmetry term of the nuclear EOS.
Comparison to BUU transport calculations indicate a somewhat asy-soft EOS.  Measurements of total nucleon emission ratios - including those
bound in clusters - is suggested as a way of compensating for possible ambiguities in coalescence calculations, which are largest at the lowest energies.   
The coalescent-invariant results
presented can be compared to transport calculations as confirmation of current theoretical understanding of the density dependence of the nuclear
asymmetry term. 


Further experimentation should also 
concentrate on observables which are predicted to be sensitive to the
high-density behavior of the asymmetry term.  This includes not only
transverse nucleon 
emission as described
in the current research, but may also include total pion production
yields \cite{li_pion_2}.  Future neutron-proton transverse emission is
planned at the NSCL 
for slightly higher 
densities, while facilities like the RI Beam Factory may be suitable
for explorations at larger $\delta$ and at densities several times $\rho_0$.
\bibliography{famiano}

\begin{thebibliography}{23}
\expandafter\ifx\csname natexlab\endcsname\relax\def\natexlab#1{#1}\fi
\expandafter\ifx\csname bibnamefont\endcsname\relax
  \def\bibnamefont#1{#1}\fi
\expandafter\ifx\csname bibfnamefont\endcsname\relax
  \def\bibfnamefont#1{#1}\fi
\expandafter\ifx\csname citenamefont\endcsname\relax
  \def\citenamefont#1{#1}\fi
\expandafter\ifx\csname url\endcsname\relax
  \def\url#1{\texttt{#1}}\fi
\expandafter\ifx\csname urlprefix\endcsname\relax\def\urlprefix{URL }\fi
\providecommand{\bibinfo}[2]{#2}
\providecommand{\eprint}[2][]{\url{#2}}

\bibitem[{\citenamefont{Lattimer and Prakash}(2001)}]{lattimer}
\bibinfo{author}{\bibfnamefont{J.}~\bibnamefont{Lattimer}} \bibnamefont{and}
  \bibinfo{author}{\bibfnamefont{M.}~\bibnamefont{Prakash}},
  \bibinfo{journal}{ApJ} \textbf{\bibinfo{volume}{550}}, \bibinfo{pages}{426}
  (\bibinfo{year}{2001}).

\bibitem[{\citenamefont{Lattimer and Prakash}((in press))}]{lattimer_bh}
\bibinfo{author}{\bibfnamefont{J.}~\bibnamefont{Lattimer}} \bibnamefont{and}
  \bibinfo{author}{\bibfnamefont{M.}~\bibnamefont{Prakash}},
  \bibinfo{journal}{Nuc. Phys. A}  (\bibinfo{year}{(in press)}).

\bibitem[{\citenamefont{Shen}(2002)}]{shen02}
\bibinfo{author}{\bibfnamefont{H.}~\bibnamefont{Shen}}, \bibinfo{journal}{Phys.
  Rev. C} \textbf{\bibinfo{volume}{65}}, \bibinfo{pages}{0535802}
  (\bibinfo{year}{2002}).

\bibitem[{\citenamefont{Page et~al.}(2005)\citenamefont{Page, Lattimer,
  Prakash, and Steiner}}]{page04}
\bibinfo{author}{\bibfnamefont{D.}~\bibnamefont{Page}},
  \bibinfo{author}{\bibfnamefont{J.}~\bibnamefont{Lattimer}},
  \bibinfo{author}{\bibfnamefont{M.}~\bibnamefont{Prakash}}, \bibnamefont{and}
  \bibinfo{author}{\bibfnamefont{A.}~\bibnamefont{Steiner}},
  \bibinfo{journal}{ApJS} \textbf{\bibinfo{volume}{593}}, \bibinfo{pages}{463}
  (\bibinfo{year}{2005}).

\bibitem[{\citenamefont{Steiner et~al.}(2005)\citenamefont{Steiner, Prakash,
  Lattimer, and Ellis}}]{steiner05}
\bibinfo{author}{\bibfnamefont{A.}~\bibnamefont{Steiner}},
  \bibinfo{author}{\bibfnamefont{M.}~\bibnamefont{Prakash}},
  \bibinfo{author}{\bibfnamefont{J.}~\bibnamefont{Lattimer}}, \bibnamefont{and}
  \bibinfo{author}{\bibfnamefont{P.}~\bibnamefont{Ellis}},
  \bibinfo{journal}{Phys. Rep.} \textbf{\bibinfo{volume}{411}},
  \bibinfo{pages}{325} (\bibinfo{year}{2005}).

\bibitem[{\citenamefont{Liu et~al.}(2004)\citenamefont{Liu, vanGoethem, Liu,
  Lynch, Shomin, Tan, Tsang, Verde, Wagner, Xi et~al.}}]{liu04}
\bibinfo{author}{\bibfnamefont{T.}~\bibnamefont{Liu}},
  \bibinfo{author}{\bibfnamefont{M.}~\bibnamefont{vanGoethem}},
  \bibinfo{author}{\bibfnamefont{X.}~\bibnamefont{Liu}},
  \bibinfo{author}{\bibfnamefont{W.}~\bibnamefont{Lynch}},
  \bibinfo{author}{\bibfnamefont{R.}~\bibnamefont{Shomin}},
  \bibinfo{author}{\bibfnamefont{W.}~\bibnamefont{Tan}},
  \bibinfo{author}{\bibfnamefont{M.}~\bibnamefont{Tsang}},
  \bibinfo{author}{\bibfnamefont{G.}~\bibnamefont{Verde}},
  \bibinfo{author}{\bibfnamefont{A.}~\bibnamefont{Wagner}},
  \bibinfo{author}{\bibfnamefont{H.}~\bibnamefont{Xi}}, \bibnamefont{et~al.},
  \bibinfo{journal}{Phys. Rev. C} \textbf{\bibinfo{volume}{69}},
  \bibinfo{pages}{014603} (\bibinfo{year}{2004}).

\bibitem[{\citenamefont{Tsang et~al.}(2004)\citenamefont{Tsang, Liu, Shi,
  Danielewicz, Gelbke, Liu, Tan, Verde, Wagner, Xu et~al.}}]{tsang_liu04}
\bibinfo{author}{\bibfnamefont{M.}~\bibnamefont{Tsang}},
  \bibinfo{author}{\bibfnamefont{T.}~\bibnamefont{Liu}},
  \bibinfo{author}{\bibfnamefont{L.}~\bibnamefont{Shi}},
  \bibinfo{author}{\bibfnamefont{P.}~\bibnamefont{Danielewicz}},
  \bibinfo{author}{\bibfnamefont{C.}~\bibnamefont{Gelbke}},
  \bibinfo{author}{\bibfnamefont{X.}~\bibnamefont{Liu}},
  \bibinfo{author}{\bibfnamefont{W.}~\bibnamefont{Tan}},
  \bibinfo{author}{\bibfnamefont{G.}~\bibnamefont{Verde}},
  \bibinfo{author}{\bibfnamefont{A.}~\bibnamefont{Wagner}},
  \bibinfo{author}{\bibfnamefont{H.}~\bibnamefont{Xu}}, \bibnamefont{et~al.},
  \bibinfo{journal}{Phys. Rev. Lett.} \textbf{\bibinfo{volume}{92}},
  \bibinfo{pages}{062701} (\bibinfo{year}{2004}).

\bibitem[{\citenamefont{Li et~al.}(1997)\citenamefont{Li, Ko, and Ren}}]{li}
\bibinfo{author}{\bibfnamefont{B.-A.} \bibnamefont{Li}},
  \bibinfo{author}{\bibfnamefont{C.}~\bibnamefont{Ko}}, \bibnamefont{and}
  \bibinfo{author}{\bibfnamefont{Z.}~\bibnamefont{Ren}},
  \bibinfo{journal}{Phys. Rev. Lett.} \textbf{\bibinfo{volume}{78}},
  \bibinfo{pages}{1644} (\bibinfo{year}{1997}).

\bibitem[{\citenamefont{Li et~al.}(1998)\citenamefont{Li, Ko, and
  Bauer}}]{parabolic}
\bibinfo{author}{\bibfnamefont{B.-A.} \bibnamefont{Li}},
  \bibinfo{author}{\bibfnamefont{C.}~\bibnamefont{Ko}}, \bibnamefont{and}
  \bibinfo{author}{\bibfnamefont{W.}~\bibnamefont{Bauer}},
  \bibinfo{journal}{Int. J. Mod. Phys. E} \textbf{\bibinfo{volume}{7}},
  \bibinfo{pages}{147} (\bibinfo{year}{1998}).

\bibitem[{\citenamefont{Zecher et~al.}(1997)\citenamefont{Zecher, Galonsky,
  Kruse, Gaff, Ottarson, Wang, Deak, Ilorvath, Kiss, Seres
  et~al.}}]{neutron_wall}
\bibinfo{author}{\bibfnamefont{P.}~\bibnamefont{Zecher}},
  \bibinfo{author}{\bibfnamefont{A.}~\bibnamefont{Galonsky}},
  \bibinfo{author}{\bibfnamefont{J.}~\bibnamefont{Kruse}},
  \bibinfo{author}{\bibfnamefont{S.}~\bibnamefont{Gaff}},
  \bibinfo{author}{\bibfnamefont{J.}~\bibnamefont{Ottarson}},
  \bibinfo{author}{\bibfnamefont{J.}~\bibnamefont{Wang}},
  \bibinfo{author}{\bibfnamefont{F.}~\bibnamefont{Deak}},
  \bibinfo{author}{\bibfnamefont{A.}~\bibnamefont{Ilorvath}},
  \bibinfo{author}{\bibfnamefont{A.}~\bibnamefont{Kiss}},
  \bibinfo{author}{\bibfnamefont{Z.}~\bibnamefont{Seres}},
  \bibnamefont{et~al.}, \bibinfo{journal}{Nuc. Instr. Meth. A}
  \textbf{\bibinfo{volume}{401}}, \bibinfo{pages}{329} (\bibinfo{year}{1997}).

\bibitem[{\citenamefont{Davin et~al.}(2001)\citenamefont{Davin, deSouza, Yanes,
  Larochelle, Alfaro, Xu, Alexander, Bastin, Beaulieu, Dorsett et~al.}}]{lassa}
\bibinfo{author}{\bibfnamefont{B.}~\bibnamefont{Davin}},
  \bibinfo{author}{\bibfnamefont{R.}~\bibnamefont{deSouza}},
  \bibinfo{author}{\bibfnamefont{R.}~\bibnamefont{Yanes}},
  \bibinfo{author}{\bibfnamefont{Y.}~\bibnamefont{Larochelle}},
  \bibinfo{author}{\bibfnamefont{R.}~\bibnamefont{Alfaro}},
  \bibinfo{author}{\bibfnamefont{H.}~\bibnamefont{Xu}},
  \bibinfo{author}{\bibfnamefont{A.}~\bibnamefont{Alexander}},
  \bibinfo{author}{\bibfnamefont{K.}~\bibnamefont{Bastin}},
  \bibinfo{author}{\bibfnamefont{L.}~\bibnamefont{Beaulieu}},
  \bibinfo{author}{\bibfnamefont{J.}~\bibnamefont{Dorsett}},
  \bibnamefont{et~al.}, \bibinfo{journal}{Nuc. Instr. Meth. A}
  \textbf{\bibinfo{volume}{473}}, \bibinfo{pages}{302} (\bibinfo{year}{2001}).

\bibitem[{\citenamefont{Phair et~al.}(1992)\citenamefont{Phair, Bowman, Gong,
  Kim, Lisa, Lynch, Peaslee, de~Souza, Tsang, and Zhu}}]{ipp}
\bibinfo{author}{\bibfnamefont{L.}~\bibnamefont{Phair}},
  \bibinfo{author}{\bibfnamefont{D.}~\bibnamefont{Bowman}},
  \bibinfo{author}{\bibfnamefont{C.~G.~W.} \bibnamefont{Gong}},
  \bibinfo{author}{\bibfnamefont{Y.}~\bibnamefont{Kim}},
  \bibinfo{author}{\bibfnamefont{M.}~\bibnamefont{Lisa}},
  \bibinfo{author}{\bibfnamefont{W.}~\bibnamefont{Lynch}},
  \bibinfo{author}{\bibfnamefont{G.}~\bibnamefont{Peaslee}},
  \bibinfo{author}{\bibfnamefont{R.}~\bibnamefont{de~Souza}},
  \bibinfo{author}{\bibfnamefont{M.}~\bibnamefont{Tsang}}, \bibnamefont{and}
  \bibinfo{author}{\bibfnamefont{F.}~\bibnamefont{Zhu}}, \bibinfo{journal}{Nuc.
  Phys. A} \textbf{\bibinfo{volume}{548}}, \bibinfo{pages}{489}
  (\bibinfo{year}{1992}).

\bibitem[{\citenamefont{Sarantites et~al.}(1996)\citenamefont{Sarantites, Hua,
  Devlin, Sobotka, Elson, Hood, LaFosse, Sarantites, and Maier}}]{microball}
\bibinfo{author}{\bibfnamefont{D.~G.} \bibnamefont{Sarantites}},
  \bibinfo{author}{\bibfnamefont{P.-F.} \bibnamefont{Hua}},
  \bibinfo{author}{\bibfnamefont{M.}~\bibnamefont{Devlin}},
  \bibinfo{author}{\bibfnamefont{L.}~\bibnamefont{Sobotka}},
  \bibinfo{author}{\bibfnamefont{J.}~\bibnamefont{Elson}},
  \bibinfo{author}{\bibfnamefont{J.}~\bibnamefont{Hood}},
  \bibinfo{author}{\bibfnamefont{D.}~\bibnamefont{LaFosse}},
  \bibinfo{author}{\bibfnamefont{J.}~\bibnamefont{Sarantites}},
  \bibnamefont{and} \bibinfo{author}{\bibfnamefont{M.}~\bibnamefont{Maier}},
  \bibinfo{journal}{Nuc. Instr. Meth. A} \textbf{\bibinfo{volume}{381}},
  \bibinfo{pages}{418} (\bibinfo{year}{1996}).

\bibitem[{\citenamefont{Ono et~al.}(1998)\citenamefont{Ono, Horiuchi, Takemoto,
  and Wada}}]{ono98}
\bibinfo{author}{\bibfnamefont{A.}~\bibnamefont{Ono}},
  \bibinfo{author}{\bibfnamefont{H.}~\bibnamefont{Horiuchi}},
  \bibinfo{author}{\bibfnamefont{H.}~\bibnamefont{Takemoto}}, \bibnamefont{and}
  \bibinfo{author}{\bibfnamefont{R.}~\bibnamefont{Wada}},
  \bibinfo{journal}{Nuc. Phys. A} \textbf{\bibinfo{volume}{630}},
  \bibinfo{pages}{148c} (\bibinfo{year}{1998}).

\bibitem[{\citenamefont{Li et~al.}(2006)\citenamefont{Li, Chen, Yon, and
  Zuo}}]{double_ratio}
\bibinfo{author}{\bibfnamefont{B.-A.} \bibnamefont{Li}},
  \bibinfo{author}{\bibfnamefont{L.-W.} \bibnamefont{Chen}},
  \bibinfo{author}{\bibfnamefont{G.-C.} \bibnamefont{Yon}}, \bibnamefont{and}
  \bibinfo{author}{\bibfnamefont{W.}~\bibnamefont{Zuo}},
  \bibinfo{journal}{Phys. Lett. B} \textbf{\bibinfo{volume}{634}},
  \bibinfo{pages}{378} (\bibinfo{year}{2006}).

\bibitem[{\citenamefont{Chen et~al.}(2004)\citenamefont{Chen, Ko, and
  Li}}]{chen04}
\bibinfo{author}{\bibfnamefont{L.-W.} \bibnamefont{Chen}},
  \bibinfo{author}{\bibfnamefont{C.}~\bibnamefont{Ko}}, \bibnamefont{and}
  \bibinfo{author}{\bibfnamefont{B.-A.} \bibnamefont{Li}},
  \bibinfo{journal}{Phys. Rev. C} \textbf{\bibinfo{volume}{69}},
  \bibinfo{pages}{054606} (\bibinfo{year}{2004}).

\bibitem[{\citenamefont{Li}(2006)}]{li_disc}
\bibinfo{author}{\bibfnamefont{B.-A.} \bibnamefont{Li}},
  \bibinfo{journal}{private communication}  (\bibinfo{year}{2006}).

\bibitem[{\citenamefont{Li and Chen}(2005)}]{li_chen05}
\bibinfo{author}{\bibfnamefont{B.-A.} \bibnamefont{Li}} \bibnamefont{and}
  \bibinfo{author}{\bibfnamefont{L.-W.} \bibnamefont{Chen}},
  \bibinfo{journal}{Phys. Rev. C} \textbf{\bibinfo{volume}{72}},
  \bibinfo{pages}{064611} (\bibinfo{year}{2005}).

\bibitem[{\citenamefont{Steiner and Li}(2005)}]{steiner05b}
\bibinfo{author}{\bibfnamefont{A.}~\bibnamefont{Steiner}} \bibnamefont{and}
  \bibinfo{author}{\bibfnamefont{B.-A.} \bibnamefont{Li}},
  \bibinfo{journal}{Phys. Rev. C} \textbf{\bibinfo{volume}{72}},
  \bibinfo{pages}{041601R} (\bibinfo{year}{2005}).

\bibitem[{\citenamefont{Lattimer and Prakash}(2004)}]{lattimer_pp}
\bibinfo{author}{\bibfnamefont{J.}~\bibnamefont{Lattimer}} \bibnamefont{and}
  \bibinfo{author}{\bibfnamefont{M.}~\bibnamefont{Prakash}},
  \bibinfo{journal}{astro-ph/0405262}  (\bibinfo{year}{2004}).

\bibitem[{\citenamefont{Slane et~al.}(2002)\citenamefont{Slane, Helfand, and
  Murray}}]{slane}
\bibinfo{author}{\bibfnamefont{P.}~\bibnamefont{Slane}},
  \bibinfo{author}{\bibfnamefont{D.}~\bibnamefont{Helfand}}, \bibnamefont{and}
  \bibinfo{author}{\bibfnamefont{S.}~\bibnamefont{Murray}},
  \bibinfo{journal}{ApJL} \textbf{\bibinfo{volume}{571}}, \bibinfo{pages}{45}
  (\bibinfo{year}{2002}).

\bibitem[{\citenamefont{Carriere et~al.}(2003)\citenamefont{Carriere, Horowitz,
  and Piekarewicz}}]{carriere}
\bibinfo{author}{\bibfnamefont{J.}~\bibnamefont{Carriere}},
  \bibinfo{author}{\bibfnamefont{C.}~\bibnamefont{Horowitz}}, \bibnamefont{and}
  \bibinfo{author}{\bibfnamefont{J.}~\bibnamefont{Piekarewicz}},
  \bibinfo{journal}{ApJ} \textbf{\bibinfo{volume}{593}}, \bibinfo{pages}{463}
  (\bibinfo{year}{2003}).

\bibitem[{\citenamefont{Li}(2002)}]{li_pion_2}
\bibinfo{author}{\bibfnamefont{B.-A.} \bibnamefont{Li}},
  \bibinfo{journal}{Phys. Rev. Lett.} \textbf{\bibinfo{volume}{88}},
  \bibinfo{pages}{192701} (\bibinfo{year}{2002}).

\end{thebibliography}
\end{document}